\documentclass[11pt,a4paper]{article}
\usepackage{amsmath,amssymb}
\usepackage{epsfig,graphicx}

\newcommand{\beq}{\begin{equation}}
\newcommand{\eeq}[1]{\label{#1}\end{equation}}
\newcommand{\bea}{\begin{eqnarray}}
\newcommand{\eea}[1]{\label{#1}\end{eqnarray}}

\begin{document}

\title{\bf Old and New No Go Theorems on Interacting Massless Particles in Flat Space\footnote{Based on the talk given at the conference {\em Quarks 2012}, Yaroslavl, Russian Federation, June 4-10, 2012.}}
\author{M. Porrati\footnote{{\bf e-mail}: massimo.porrati@nyu.edu}
\\
\small{\em CCPP, Department of Physics, NYU} \\
\small{\em 4 Washington Pl. New York NY 10016 (USA)}
}
\date{}
\maketitle

\begin{abstract}
We review model independent arguments showing 
that massless particles interacting with
gravity in a Minkowski background space can have at most spin two. These 
arguments include a classic theorem due to Weinberg, as well as a more recent extension of the Weinberg-Witten theorem. A puzzle arising from an apparent counterexample to these theorems is examined and resolved.
\end{abstract}

\section{Introduction}
Interacting theories of high spin particles are notoriously fraught with 
problems. In particular, when 
particles are massless  and spacetime is Minkowski, there seem to exist obstacles to the very existence of consistent interactions~\cite{D}. My contribution to the proceeding of this conference shall review old and new {\em no go} theorems on interacting high-spin theories. These theorems are surprisingly strong when combined together and rule out {\em any} interaction between particles of spin higher than 2 and any matter that interacts with gravity. The same theorems also show that
there exists only one massless spin two particle interacting with gravity, i.e. the graviton itself.~\footnote{These theorems assume a mild form of locality that in our view is indispensable to ensure that a theory is unitary and causal. By relaxing these requirements, interacting theories seem to exist~\cite{T}. We shall briefly discuss later their
potential problems.}

\section{Old and New  {\em No Go} Theorems}
\subsection*{Weinberg 1964}
An important obstruction to consistent interactions of high-spin massless 
particles was derived in 1964 by Weinberg~\cite{w64} (see also~\cite{w65}) using general properties
of the S-matrix. His result was extended to Fermions and specifically to
supersymmetric theories in~\cite{gp,gpv}. 

Weinberg considers an S-matrix element with $N$ external particles of four-momentum 
$p_i$, 
$i=1,..N$ and one massless spin-$s$ particle of momentum $q$ and polarization 
vector $\epsilon^{m_1..m_s}(q)$. In the soft limit
$q\rightarrow 0$, the interaction  of a single  spin-$s$ particle is described by a matrix element of a spin-$s$ current $J_{m_1.... m_s}$. Moreover, since in the soft limit the scattering 
of the particle is elastic, the matrix element obeys
\beq
\lim_{q\rightarrow 0}\langle p A|J_{mnr...}|A' p'\rangle = g\delta_{AA'} p_mp_np_r...
\eeq{q1}
Here $A,A'$ denote all quantum numbers other than momentum for the initial/final hard particle --i.e. the particle that either absorbs or emits the soft, spin-$s$ particle.
When the spin-$s$ particle ends on an external line, the S-matrix exhibits a pole 
$1/pq$. So, up to regular terms  it factorizes as (see fig. 1)
\beq
S(p_1,..,p_N,q,\epsilon)\approx \sum_{i=1}^N g^i{p^i_{m_1}...p^i_{m_s}
\epsilon^{m_1..m_s}(q) \over 2p^i q} S(p_1... p_N).
\eeq{2}

\begin{figure}[h]
\includegraphics[width=15.9cm]{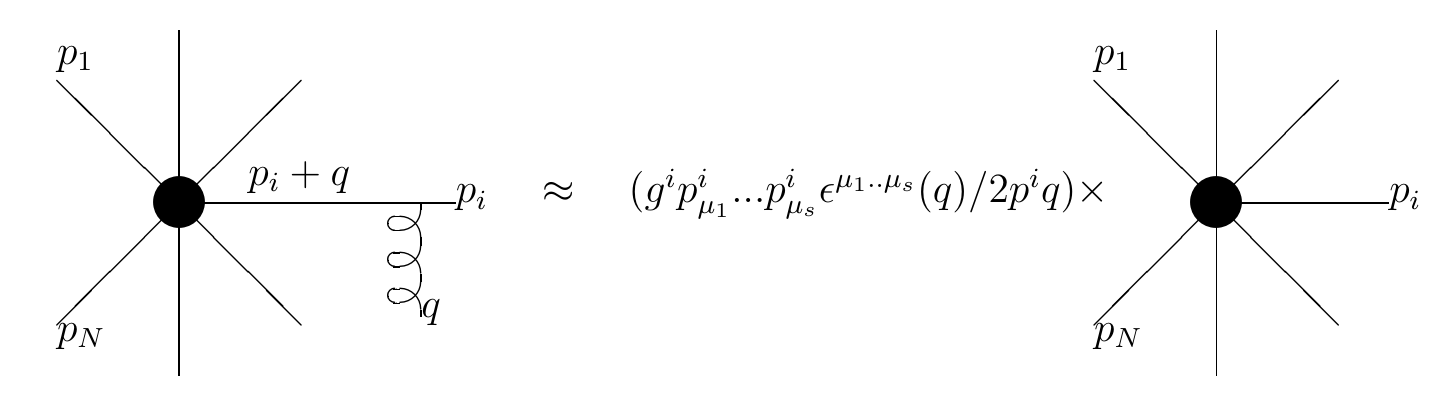}
\caption{Factorization of S-matrix amplitude in the soft limit}
\label{fig1}
\end{figure}
\vskip .1in
\noindent
The polarization vector is transverse and traceless, so it gives a redundant 
description of the massless particle, which has only two physical 
polarizations. Redundancy is eliminated by demanding that the S-matrix be
independent of spurious polarizations
\beq
\epsilon_{spurious}^{m_1..m_s}(q)\equiv q^{(m_1}\eta^{m_2..m_s)}(q),
\qquad q_n\eta^{nm_1..m_{s-2}}(q)=\eta_n^{nm_1..m_{s-3}}(q)=0.
\eeq{4}
The factorization eq.~(\ref{2}) implies that spurious 
polarizations decouple only when
\beq
\sum_i g^i p^i_{m_1}...p^i_{m_{s-1}}=0, \qquad \forall p_i.
\eeq{5}
For generic momenta this equation has a nonzero solution only in two cases:
\begin{description}
\item{$s=1$} In this case eq.~(\ref{5}) reduces to $\sum_i g^i=0$, i.e. to 
conservation of charge.
\item{$s=2$} Here eq.~(\ref{5}) becomes $\sum_i p_m^i=0$ and $g^i=\kappa$. 
The first equation enforces energy-momentum conservation, 
while the second gives the 
principle of equivalence: all particles must interact with the massless spin
two with equal strength $\kappa$.
\end{description}
For $s>2$ eq.~(\ref{5}) has no solution for generic momenta.

This argument shows that only scalars, vectors and {\em one} spin two particle 
can give rise to 
long-distance interactions. There can exist only one graviton with long range interactions 
because if there were two, then Weinberg's theorem would say that the first 
would interact with all 
matter with a universal constant, say $g$, while the second would interact with a universal 
constant $g'$. Now, if one calls $J_{mn}$ the current associated with the first graviton and 
$J'_{mn}$  that associated with the second, then the current 
$g' J_{mn}-gJ_{mn}$ describes a ``non-graviton," which does not produce 
long range interactions with matter. The real graviton is associated with the current $gJ_{mn}+g'J_{mn}$. This argument extends trivially to $N>1$ massless spin-two particles.

Weinberg's theorem was extended  to Fermions 
in~\cite{gp,gpv}. There, it was shown that massless 
Fermions only
up to spin 3/2 interact at low 
energies\footnote{Spin 3/2 Fermions were also shown to interact as the
supersymmetric partners of the graviton, i.e. the gravitini of 
supergravity theory.}.
Both~\cite{w64} and~\cite{gp,gpv} rely on the existence of processes 
in which the number
of spin-$s$ particles changes by one unit. This is necessary to generate 
long-range interactions, but it leaves out the possibility of
interacting high-spin particles with a nonzero conserved charge. 
In particular, particles interacting only with the graviton according to the 
principle of equivalence are still allowed. 

One may think that another way out of Weinberg's theorem is to soften the interaction of $s>2$ particles by appropriate powers of the soft momentum 
$q_m$ in such a manner as to cancel the offending pole in eq.~(\ref{2}). If this were true, high-spin particles would not generate long range forces, but they could still interact..... {\em but not even this is true}. When combined with another, more recent no go theorem~\cite{p08}, Weinberg's theorem {\em completely} excludes interactions between $s>2$ particles and any object that interacts with gravity --and also proves that only one interacting massless graviton exists.

The no go theorem of ref.~\cite{p08} is an extension of the famous Weinberg Witten theorem~\cite{ww}. It can be obtained by studying the matrix element 
\beq
\lim_{p'\rightarrow p} \langle \epsilon p| T_{mn} | \epsilon' p' \rangle,
\eeq{q2}
which describes the absorption (or emission) of a soft graviton by an on-shell  massless spin-$s$ particle, 
described by the polarization tensors $\epsilon,\epsilon'$, with initial momentum $p$ and final momentum $p'$. Since the particle is on shell before and after interacting with the graviton, the graviton momentum $q=p'-p$ is off-shell and space-like. 

When the polarization tensor $\epsilon'=\epsilon$ describes a physical state, one must have
\beq
\lim_{p'\rightarrow p} \langle \epsilon p| T_{mn} | \epsilon p' \rangle =p_mp_n  ,
\eeq{q3}
This is a special case of Eq.~(\ref{q1}). The coupling constant $g=1$ is the same for all 
states because of Weinberg's theorem.  
 
\subsection*{The Weinberg Witten Theorem and Beyond}
When the polarization tensor $\epsilon'$ describes a spurious state, the matrix element~(\ref{q2}) must not contribute to a physical scattering amplitude. One obvious way to achieve this is for the matrix element itself to vanish
\beq
\lim_{p'\rightarrow p} \langle \epsilon p| T_{mn} | \epsilon' p' \rangle =0\qquad \mbox{when $\epsilon'$ is spurious}.
\eeq{q4}
As shown in~\cite{ww}, this condition implies that the spin-$s$ massless particles can have at most spin one. We refer to that reference for an elegant 
proof of this statement.

Eq.~(\ref{q4}) guarantees decoupling of spurious states, but it can be weakened. The reason is that the graviton in~(\ref{q2}) is off-shell and so it contributes a factor 
$\Pi^{mn}_{pq}(q)$ (the graviton propagator) to the scattering amplitude. 
If the matrix element is proportional to the graviton linearized equations of 
motion~\footnote{Schematically denoted here by 
$L_{mn}^{pq}h_{pq}=0$, with $h_{mn}$ the metric fluctuation around Minkowski space.} as in
\beq
\langle \epsilon p| T_{mn} | \epsilon' p'\rangle \propto L_{mn}^{pq}(p-p')\Delta_{pq}(q),
\eeq{q5}
then it contributes to the amplitude a factor 
\beq
 \langle \epsilon p| T_{mn} | \epsilon' p'\rangle \Pi^{mn}_{pq}(q)\propto \Pi^{mn}_{pq}
L_{mn}^{rs}(q)\Delta_{rs}(q) =\Delta_{pq}(q).
 \eeq{q6}
If $\Delta_{pq}(q)$ is analytic near $q=0$, the spurious states contribution {\em may} 
be canceled by a contact term~\footnote{In Field theory language this is equivalent to say that the change in the action due to the linear gauge transformation of the spin-$s$ field can be canceled by a change in the graviton field.}. So, eq~(\ref{q5}) is the most general {\em necessary} condition for consistency of a high-spin massless particle interacting with gravity. 
 
 Ref.~\cite{p08} gives a prof of this statement, here we shall only sketch it out (for Fermions, when it is technically a bit simpler).
 
The matrix element $\langle \epsilon, p+q | T_{mn} | \epsilon', p \rangle $ is bilinear
in $\epsilon,\epsilon'$ and it otherwise depends only on the momenta. For spin $s$, 
the minimum set of
spurious states needed to write a nonzero conserved, symmetric tensor is
given by Dirac spinor-tensors $\epsilon_{\alpha,\; m_1...m_n}(p)$, $s=n+1/2$. 
They are symmetric in the vector indices $m_1,..,m_n$ and satisfy the
constraints  
\beq
/\!\!\!p \epsilon_{m_1,..,m_n}(p)=0, \qquad p^{m_1} \epsilon_{m_1,..,m_n}(p)=0,
\qquad \gamma^{m_1} v_{m_1,..,m_n}(p).
\eeq{19}
We are interested in initial and final states with the same physical helicity 
$+s$, so we impose on the representatives of the initial state ($\epsilon$) and final state 
($\epsilon'$) the equations 
\beq
\gamma^5 \epsilon_{m_1,..,m_n}(p) =\epsilon_{m_1,..,m_n}(p), \qquad
\gamma^5 \epsilon'_{m_1,..,m_n}(p+q) =\epsilon'_{m_1,..,m_n}(p+q).
\eeq{20}

In the kinematic configuration of interest, there exist two independent 
light-like vectors: $p$ and $p+q$. The space-like vector $q$ can be used to
define $n+1$ algebraically independent spinor-tensors
\beq
\epsilon^k_{m_1,..,m_k}(p)\equiv q^{m_{k+1}}...q^{m_n} \epsilon_{m_1,..,m_n}(p),
\qquad k=0,..,n.
\eeq{21}

Constraints~(\ref{19},\ref{20}) and the on-shell condition on momenta, $p^2=
(p+q)^2=0$, vastly reduce the possible terms in the 
matrix element of interest. 
A short reflection suffices to convince oneself that its most general form is
\beq
\langle \epsilon, p+q | T_{mn} | \epsilon', p \rangle = \sum_{k=0}^n 
A^k \bar{\epsilon}^k (p+\alpha^k q)_{(m} \gamma_{n)} \epsilon'^k + \sum_{k=1}^n B^k \bar{\epsilon}^k_{(m}\gamma^{}_{n)} \epsilon'^{k-1} + \sum_{k=1}^n C^k \bar{\epsilon}^{k-1}_{(m}\gamma^{}_{n)} \epsilon'^{k}.
\eeq{25}
The coefficients $A^k$ ,$B^k$, $C^k$ and $\alpha^k$ are functions of $q^2$ 
which in principle can be singular at $q^2=0$. A first constraint on the
singularity is due to the principle of equivalence, that demands
\beq
 \lim_{q\rightarrow 0} \langle \epsilon, p+q | T_{mn} | \epsilon', p \rangle =
p_m p_n.
\eeq{26}
This equation implies
\bea
\lim_{q\rightarrow 0} A^n(q) &= & 1,\label{27} \\
\lim_{q\rightarrow 0} A^{k}(q)q^{2(n-k)}&=& 0, \qquad k<n ,\label{28}\\
\lim_{q\rightarrow 0} \alpha^k(q) A^k(q) q^{2(n-k)-1} &=& 0 , \label{29} \\
\lim_{q\rightarrow 0} B^k(q) q^{2(n-k)+1}&=&0 , \label{30}\\
\lim_{q\rightarrow 0} C^k(q) q^{2(n-k)+1}&=& 0 . 
\eea{31}
Conservation of $T_{mn}$ implies that the matrix element~(\ref{25}) is
divergenceless 
\beq
q^m \langle \epsilon, p+q | T_{mn} | \epsilon', p \rangle =0. 
\eeq{31a}
This
yields the further constraints
\beq
A^k (\alpha^k -1/2) q^2 + B^{k+1} + C^{k+1} = 0, \qquad k=0,.., n-1 , \qquad
\lim_{q\rightarrow 0}\alpha^n (q) =1/2.
\eeq{33}
Though not strictly necessary to prove our result, eq.~(\ref{33}) is useful
since it simplifies the structure of the matrix element. In particular, 
together with the mass-shell conditions~(\ref{19}) it makes the matrix element 
transverse and traceless. 

In reality, constraints~(\ref{27}-\ref{31}) are too weak, because if any of 
the  coefficients 
$A^k$ ,$B^k$, $C^k$ and $\alpha^k A^k$ had a singularity $1/q^2$~\footnote{For
instance $A^k(q)=A_r^k(q)q^{-2}$, $A^k_r(q)=$ regular and nonzero at $q^2=0$.}
then vertex~(\ref{25}) would imply the existence of another massless spin 2
field (it couples to a transverse-traceless vertex!) which mixes linearly with
the graviton. This linear mixing contradicts Weinberg's 
uniqueness theorem. 
It also violates the principle of equivalence --which we
assumed (and need) to prove or theorem-- either because it
implies the existence of a second massless graviton that couples only to some
type of matter, or because it re-sums to give the 
graviton a
mass.  A singularity stronger than $1/q^2$ is even worse since it implies the
existence of a spin two ghost mixing linearly with the ordinary graviton 
(see fig. 2).
\begin{figure}[h]
\includegraphics[width=10cm]{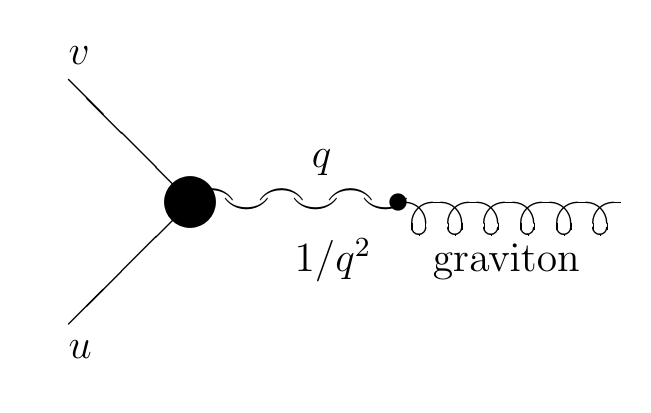}
\caption{A singular vertex implies the existence of an additional massless
particle mixing with the graviton.}
\label{fig2}
\end{figure}
\vskip .1in
\noindent
We have introduced extra polarizations to ensure that the $T_{mn}$ matrix
element transforms covariantly. Now we must check under which conditions 
spurious polarizations do indeed decouple. Spurious states have the form
\beq
\epsilon'_{s\;m_1...m_n}(p)=p^{}_{(m_1}\eta_{m_2.... m_n)},
\eeq{34}
where $\eta_{m_1.... m_{n-1}}$ is on shell, transverse and 
gamma-transverse. For the spurious state~(\ref{34}), 
the spinor-tensors given in eq.~(\ref{21}) have the form
\beq
\epsilon'^{k}_{s\;m_1...m_k}(p)= p_{(m_1}\eta^{k-1}_{m_2.... m_k)} 
-(n-k){q^2 \over 2} \eta^{k}_{m_1.... m_k} ,
\qquad 
\eta^{k}_{m_1.... m_k} \equiv q^{m_{k+1}}...q^{m_n}
\eta^{k}_{m_1.... m_n}.
\eeq{35}
Matrix element~(\ref{25}) is transverse and traceless, therefore the decoupling
condition~(\ref{q5}) simplifies to 
\beq
\langle \epsilon, p+q | T_{mn} | \epsilon'_s, p \rangle = q^2 \Delta_{mn}(q).
\eeq{36}
Substitution of eqs.~(\ref{34},\ref{35}) into eq.~(\ref{25}) then
yields a set of 
recursion relations among the coefficients $A^k,...,C^k$:
\bea
-kA^k -{q^2\over 2}(n+1-k)A^{k-1} +C^k &=& {\cal O}(q^2), \qquad
k=1,...,n;\label{37}\\
-k\alpha^k A^k -{q^2\over 2}(n+1-k)\alpha^k A^{k-1} &=& {\cal O}(q^2), 
\qquad k=1,...,n;
\label{38}\\
-(k-1)B^k -{q^2\over 2}(k+2-k)B^{k-1} &=& {\cal O}(q^2), \qquad k=2,...,n;
\label{39}\\
-(k-1)C^k -{q^2\over 2}(k+1-k)C^{k-1} &=& {\cal O}(q^2), \qquad k=2,...,n.
\eea{40}
As we have seen earlier, no coefficient in eq.~(\ref{25}) can be more 
singular than $1/q^2$. So in particular
\beq
\lim_{q\rightarrow 0}q^2 C^1(q)=\lim_{q\rightarrow 0}q^2 A^0(q)=0.
\eeq{41}
Recursion relations~(\ref{37}, \ref{40}) then imply
\beq
\lim_{q\rightarrow 0}A^n(q)=0, \qquad n>1,
\eeq{42}
in contradiction with the equivalence principle, which implies 
$A^n(0)=1$ [see eq.~(\ref{27})].

This completes our proof: only when spurious polarizations decouple from the
cubic vertex~(\ref{25}) a chance exists for massless high-spin fields to
interact with gravity, but decoupling contradicts the universality of
gravitational interactions. 

Our argument rules out interactions for Fermions of spin $s> 3/2$. It still
allows for gravitational interactions of  spin 3/2 particles. This is not
surprising since supergravity theories provide many examples of massless
spin 3/2 particles consistently interacting with gravity and other fields.

When the argument is repeated for Bosons, it shows that particles of spin 
higher than $2$ cannot interact with gravity. 

\subsubsection*{Combining Old and New Theorems}
\begin{figure}[h]
\includegraphics[width=15.9cm]{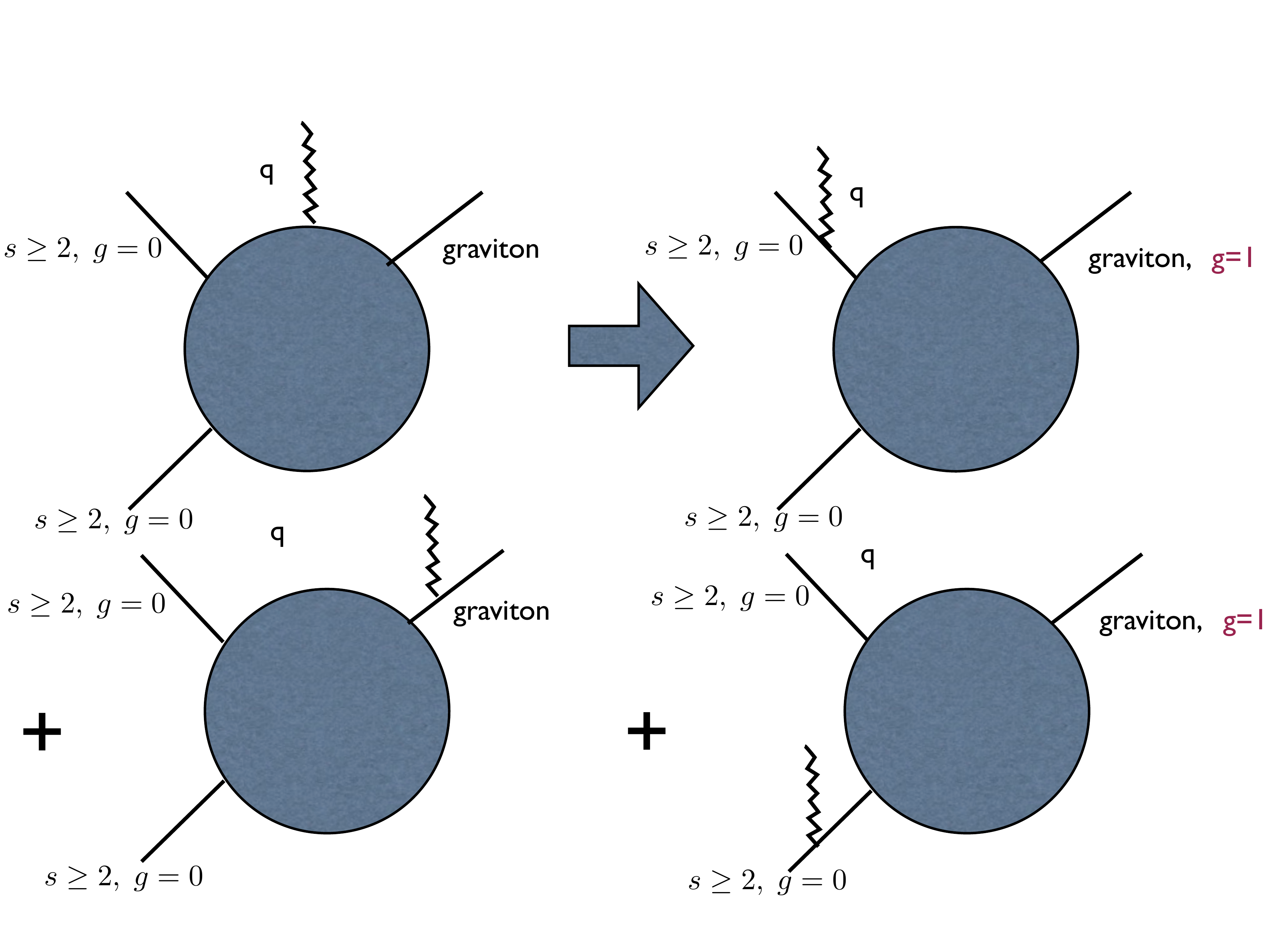}
\caption{``Exotic" particles cannot interact {\em at all} with gravity.}
\label{fig3}
\end{figure}

The ``new" theorem just described here does not {\em yet}  
forbid the existence of some exotic matter, which does not interact with gravity according to the 
principle of equivalence. But such possibility is ruled out by combining our ``new" theorem 
with Weinberg's ``old" one. Consider indeed an amplitude with one soft
graviton and three particles with {\em arbitrary} momentum: two ``exotic" particles of spin 
$s\geq 2$, and one graviton. It factorizes as in Eq.~(\ref{2})  [see figure~(\ref{fig3})].
The soft graviton couples with exotic matter with coupling constant $g=0$ because of
our new theorem, but it couples with the graviton with $g=1$. This contradicts Weinberg's old theorem, so the amplitude must vanish. The argument can be repeated with $N$ ``exotic" and $M$ ``standard" particles, and the result is similar: the amplitude factorizes into one containing only exotic particles times one containing
only normal ones. So, exotic particles cannot interact at all with standard ones. The result can be summarized in field theory language as follows: call $\psi$ the exotic fields and $\phi$ the standard ones (including the graviton); then, the above argument states that the action factorizes completely as
\beq
S[\phi,\psi]=S_1[\phi]+ S_2[\psi]
\eeq{q7}
Notice that the action of the exotic particles, $S_2[\psi]$, does not depend on {\em any} standard field,  not even on the metric. So, by any meaningful definition of the word ``exist," exotic particles do not.

\section{A Counterexample that Isn't}
An excellent way to test the robustness of a no go theorem is to search for possible 
counterexamples. Here we shall examine an apparent one and explain why in reality it does not
contradict our theorem~\cite{fp}. 

In Anti de Sitter spaces it is possible to construct interacting theories of massless high spin states. These theories were pioneered by Fradkin and Vasiliev in~\cite{fv}.  Modern reviews on this vast subject are e.g.~\cite{v,review}; 
their relations to and ways out of no go theorems have been reviewd 
in~\cite{X}~\footnote{Vasiliev's theories ``evade'' no go theorems because 
such theories are defined in Anti de Sitter space instead of Minkowski space.
The cosmological constant of AdS introduces a new mass scale that allows for
Lagrangian interactions, {\em singular in the flat space limit}, 
that decouple spurious polarizations.}. A puzzle arises now if we cut off $AdS_5$ space in a Vasiliev theory; in other words, if we extend the  Randall Sundrum 2 (RS2)~\cite{rs2} construction to high-spin theories. 

Even for Vasiliev theories, a crucial  property of the Randall-Sundrum construction holds: 
all states with $k^2=0$ and a normalizable 5D wave function appear as massless particles in 4D Minkowski space. 
These particles interact with each others and with the graviton. Moreover, they are invariant under all the spin-$s$ gauge symmetries needed to decouple spurious polarizations, because this is the defining property of the Fradkin-Vasiliev construction~\cite{fv}. Finally, their interactions with the graviton obeys the principle of equivalence~\footnote{This is seen best by using
the duality between Vasiliev's theory and free three-dimensional $O(N)$ models 
conjectured in~\cite{kp} and
further elaborated and tested in~\cite{gy}. In the 3D picture, the principle of equivalence
for high-spin particles follows from the fact that correlators of high-spin currents 
with $T_{mn}$ obey standard Ward identities (see also~\cite{mz}).}. But this contradicts all 
our no go theorems! So, something in the previous line of reasoning must be incorrect. One possibility is that high-spin fields do not give normalizable modes at $k^2=0$. This possibility is excluded by computing the induced 4D kinetic term. Explicitly, for a spin-$s$ field, $\psi_{m_1,..m_s}(z,x)$, $m_i=0,1,2,3,4$, the $k^2=0$ wave function must behave as $\psi_{m_1,..m_s}(z,x)=z^E \hat{\psi}_{m_1,..m_s}(x)$  and the $m_i=4$ components can be set to zero with a gauge choice. Since $\psi_{\mu_1,..\mu_s}(x)$ is a rank-$s$ covariant tensor~\footnote{In 
this subsection, 4D indices are denoted by Greek lowercase letters while Latin
lowercase letters denote 5D indices.}, 
the relation between $E$ and the scaling dimension $\Delta$ is 
$E=4-\Delta -s$~\cite{fp}. 
Thus, the 5D kinetic term of the massless mode for rank-$s$ field becomes
\bea
\int d^4x dz \sqrt{-g} g^{m_1 n_1}.... g^{m_{s+1} n_{s+1}} \partial_{m_1}\psi_{m_2... m_{s+1}} (z,x)\partial_{n_1} \psi_{m_2... m_{s+1}}(z,x) +.... & & \nonumber \\
\propto  \int_\epsilon^\infty dz z^{2s-3}z^{8-2s-2\Delta}\int d^4x\partial_{\mu_1}\hat{\psi}_{\mu_2... \mu_{s+1}} (x)\partial^{\mu_1}\hat{\psi}^{\mu_2... \mu_{s+1}} (x) +.....&&
\eea{m12}
The dimension $\Delta$ is $s+2$ thus, the integral in $dz$ in eq.~(\ref{m12}) converges for all spins $s>1$. 

The problem arises with interactions, as it may have been expected; specifically with the spin-$s$ gauge invariance of interactions, starting at the first nontrivial order (i.e. cubic order in the fields). When we stated that the Fradkin-Vasiliev construction ensured gauge invariance of the full interacting action under spin-$s$ gauge transformations, we implicitly assumed that the gauge 
transformations were normalizable near $z=0$. This is the correct boundary condition for $AdS_5$, but not for $AdS_5$
cut off at $z=\epsilon$. In the latter case, it is precisely the (cutoff) non-normalizable gauge transformations that become the gauge
transformations of 4D massless particles. It is they that ensure the 
decoupling of
4D spurious polarizations. It is also they that may not leave the action invariant, because for them it is no longer legitimate to integrate by part and discard boundary terms. 

To be specific, let us start by writing down the inhomogeneous part of the spin-$s$ gauge transformation
\beq
\delta\psi_{m_1,..m_s}(z,x)=D_{(m_1}\epsilon_{m_2... m_s)_T} (z,x).
\eeq{m13}
The 4D gauge transformations are those that leave the field $\psi_{m_1,..m_s}(z,x)=z^E \hat{\psi}_{m_1,..m_s}(x)$ in the
gauge $\psi_{4, m_2..m_s}=0$. This condition constrains the gauge parameter to have the form
\beq
\epsilon_{(m_1... m_{s-1})_T}=z^{2-2s}\hat{\epsilon}_{(m_1... m_{s-1})_T}. 
\eeq{m14}

Next, decompose the bulk action into a free quadratic part plus an interacting part. The quadratic part is, schematically,
\beq
S_2=\int d^4x dz \psi(z,x)(L \psi)(z,x) + \int d^4x \hat{\psi}(x) (B\hat{\psi})(x).
\eeq{m15}
The kinetic term of the bulk quadratic action has been denoted here by $L$, while all boundary terms needed to enforce Neumann boundary conditions have been called $B$. These terms ensure that when $\hat{\psi}$ obeys the 4D equations of motion of a free massless spin-s particle, then the action is stationary {\em even under variation that do not vanish at the boundary} $z=\epsilon$, such as those given by eq.~(\ref{m13}) with gauge parameter~(\ref{m14}).

Interactions arise first at cubic order. One universal interaction term that is always present involves two spin-$s$ fields and a metric fluctuation $h_{mn}(z,x)\equiv L^{-2} z^{2} g_{mn} -\eta_{mn}$. Schematically the action is
\beq
S=S_2+S_3, \qquad S_3 = \int d^4x dz V[\psi(x),\psi(x),h(x)].
\eeq{m16}
The local cubic interaction $V[\psi(x),\psi(x),h(x)]$ is a sum of two terms: $V_m+V_{FV}$. The first one comes from covariantizing the 
action~(\ref{m15}) using the minimal coupling procedure and expanding to linear order in $h_{mn}$. The second is the Fradkin-Vasiliev (FV) vertex~\cite{fv}, which is needed to ensure consistency of the equations of motion to cubic order (see~\cite{bl,bls} for a recent derivation of this vertex). 

Now a crucial observation is that if the fields $\psi$ are on shell, i.e. if they obey both the 5D bulk equations as well as the 4D equations of motion, then the quadratic action is invariant under arbitrary variations $\delta \psi$. 
So, under a full non-linear gauge variation and up to quadratic  order in the fields, the change in $S_2+S_3$ reduces to the change in $S_3$:
\beq
\delta S = \int d^4x dz \left\{V[\delta\psi(x),\psi(x),h(x)] + V[\psi(x),\delta \psi(x),h(x)]\right\},
\eeq{m17}
with $\delta\psi$ given in eq.~(\ref{m13}).

To find out where the problem lays with the would be massless high-spin states,
 we consider now the explicit case of spin 3 particles. 

The cubic FV vertex for two $s=3$ particles and a graviton can be found e.g. 
in~\cite{bls} eq. (14). It simplifies dramatically in the gauge $\psi_{4 mn}(z,x)=0$, $h_{4m}(z,x)=0$, especially when the $s=3$ field $\psi_{\mu\nu\rho}$ is on shell and  the metric fluctuation $h_{\mu\nu}$ is independent of $z$. These are the field configurations we need to show the problem with interactions. On such configuration  the FV vertex contains only one term involving derivatives w.r.t. $z$ [compare with~\cite{bls} eq. (14)]:
\beq
\int d^4x dzV_{FV}[\psi,\psi,h]= -{3\over \Lambda} \int d^5x dz {L^5\over z^5} w_{\alpha\beta\gamma\delta}D^z \psi^{\mu\alpha\beta}D_z\psi^{\gamma\delta}_\mu. 
\eeq{m18}
Here $\Lambda$ and $w^{\alpha\beta\gamma\delta}$ are, respectively,  the 5D cosmological constant and the linearized Weyl tensor.
Inserting the gauge variation~(\ref{m13}) into eq.~(\ref{m18}), the only way to cancel the resulting
 term is to integrate by part in $dz$. Integration
by part produces a term that cancels against lower-dimension terms upon using 
the free $\psi$ equations of motion. But it also produces the  boundary term
\beq
-{3\over \Lambda}\int d^4x {L^5\over \epsilon^5} w_{\alpha\beta\gamma\delta}D^{(\mu}\epsilon^{\alpha\beta)_T}
\stackrel{\leftrightarrow}{D}_z\psi^{\gamma\delta}_\mu .
\eeq{m19}
This term would have been zero on a metrically complete $AdS_5$ space and for a normalizable gauge variation. On cutoff $AdS_5$ instead, it does not vanish. Moreover, since the only field that is not on shell with respect to the 4D equations of motion is the metric fluctuation, the only chance to cancel~(\ref{m19}) is
by a {\em local} variation of $h_{\mu\nu}$. For this to be possible, eq.~(\ref{m19}) would have to  vanish when
$h_{\mu\nu}$ is on shell. But eq.~(\ref{m19}) is proportional to the Weyl tensor, which does not vanish on shell!

{\em Therefore, the gauge symmetry~(\ref{m13}) is anomalous to first order in the gravitational interactions}.

This is in exact agreement with our no go theorems. As for any anomalous gauge symmetry, the only way to
escape algebraic inconsistency is for the high spin field to acquire a mass.

Luckily, a mass counterterm is natural in the RS2 construction, since it can be introduced simply by modifying the term 
$B$ in eq.~(\ref{m15}). So, a RS2-type construction is possible even for Vasiliev high-spin theories but the resulting  dynamical gravity plus matter in 4D Minkowski space  does not contain massless particles of spin higher than 2.
\section{End Note on Locality}
Recently, a  theory of massless high spin theory has been proposed 
in~\cite{T}, that evades our theorems. It does so by having quartic Lagrangian vertices that contain non-local terms $\propto 1/pq$, which cancel the spurious polarization contribution to physical S-matrix amplitudes. Because of this non-locality it seems difficult to reconcile this theory with {\em both} unitarity and causality~(see e.g. \cite{ahddg}).
\subsection*{Acknowledgments}
Work supported in part by NSF grant PHY-0758032, and
by ERC Advanced Investigator Grant n.226455 {\em Supersymmetry,
Quantum Gravity and Gauge Fields (Superfields)}.

\end{document}